\documentclass[aps,showpacs,superscriptaddress]{revtex4}

\usepackage{graphicx}

\begin{document}

\title{Survey Propagation as local equilibrium equations}

\author{Alfredo Braunstein}
\email[]{abraunst@sissa.it}
\affiliation{SISSA, Via Beirut 9, 34100 Trieste, Italy} 
\affiliation{ICTP, Strada Costiera 11, I-34100 Trieste, Italy}

\author{Riccardo Zecchina}
\email[]{zecchina@ictp.trieste.it}
\affiliation{ICTP, Strada Costiera 11, I-34100 Trieste, Italy}

\begin{abstract}
It has been shown experimentally that a decimation algorithm based on
Survey Propagation (SP) equations allows to solve efficiently some
combinatorial problems over random graphs.  We show that these
equations can be derived as sum-product equations for the computation
of marginals in an extended space where the variables are allowed to
take an additional value -- $*$ -- when they are not forced by the
combinatorial constraints.  An appropriate ``local equilibrium
condition'' cost/energy function is introduced and its entropy is
shown to coincide with the expected logarithm of the number of
clusters of solutions as computed by SP. These results may help to
clarify the geometrical notion of clusters assumed by SP for the
random K-SAT or random graph coloring (where it is conjectured to be
exact) and helps to explain which kind of clustering operation or
approximation is enforced in general/small sized models in which it is
known to be inexact.

\end{abstract}


\maketitle

\section{Introduction}

Recent developments in statistical physics of disordered systems have
shown a remarkable convergence of themes with other disciplines such
as computer science (e.g combinatorial optimization~\cite{TCS}),
information theory (e.g error correcting codes~\cite{Codes}) and
discrete mathematics (e.g. random
structures~\cite{Aldous,Guerra:Talagrand,Aldous_z2}).  While the study
of a typical static measure characterizing the slow dynamics of both
physical and algorithmic processes is the unifying issue in
out-of-equilibrium problems, the study of the geometrical structure of
ground states of spin-glass-like energy functions $E$ is central to
the understanding of the onset of computational complexity in random
combinatorial problems.  The combinatorial problem of satisfying a
given set of constraints is viewed in the physics framework as the
problem of minimizing $E$ and ``ground state configurations'',
``solutions'' or ``satisfying assignments'' should be understood as
synonymous.

Important in an attempt of providing a complete theory of random
combinatorial problems is the notion of pure states, or clusters of
configurations, on which the probability measure over optimal
configurations is assumed to concentrate.  Recently, a new class of
algorithms has been proposed \cite{MZ,science,BMZ} that have shown
surprising capabilities in dealing with the (exponential)
proliferation of clusters of metastable states and therefore in
solving random instances of combinatorial problems which are difficult
to solve for local search heuristics.  Such algorithms are based on
the so called Survey Propagation (SP) equations in which indeed a
decomposition of the ground states probability distribution -- the
Gibbs measure -- into an exponential number of clusters is assumed
from the beginning.  The SP equations can be viewed as zero
temperature cavity equations~\cite{cavity} formulated for single
instances at a level equivalent to the one-step of replica symmetry
breaking (1-RSB) scenario~\cite{states}.

The SP algorithm consists in a message-passing technique which is
closely related to another message-passing method -- known as
sum-product or Belief Propagation (BP)~\cite{Gallager,Pearl} algorithm
-- which have shown amazing performance for solving the decoding
problem~\cite{Spielman} in error correcting codes based on sparse
graph
encodings~\cite{Sourlas,turbo,Forney,good_codes1,good_codes2,MacKay}.

The aim of this study is to discuss the precise (finite size)
structure of the SP equations, linking them to the BP formalism. This
is a well defined mathematical issue, independent on the physical
origin of the equations. Due to the algorithmic relevance of both BP
and SP for coding theory and combinatorial optimization, it is a basic
question to understand what these equations are doing for a finite
number of variables $N$ since this is the regime in which they are
used.

As we shall see, the SP ``algorithmic'' equations at finite N are
performing a very specific clustering operation over the solution
space.  Moreover, the number of such clusters in the Bethe
approximation will be shown to coincide with the prediction of the
cavity theory.

These results will be obtained by showing that the SP equations are
the BP equations for a modified combinatorial problem. By this mapping
we clarify how the hypothesis making BP exact (that is, uncorrelation
of distant variables) translate onto a condition of uncorrelation of
"frozen"  variables belonging to different clusters: SP produces a
collapse of the internal structure of clusters and eliminates
correlations among the unfrozen parts.

We shall present the results in the case of the K-SAT problem even
though the method could be applied to any discrete combinatorial model
defined over locally tree--like graphs.  The results concerning the
cluster entropy will be compared with the prediction of the 1-RSB
cavity analysis for random K-SAT.

The line of reasoning of the paper consists in showing that the SP
equations can be re-derived as sum-product or BP equations --
i.e. simple replica symmetric (RS) cavity equations -- over an
extended configuration space.  The definition of this space consists
in associating to each binary variable a new extra value ``$*$'' which
will correspond to the possibility that the variable is not forced to
take one of the binary values $\{ -1,+1 \}$ in a given solution
\cite{pspin-note}.  We will introduce a {\it local equilibrium
condition} (LEC) cost-energy function $\hat E$ derived from $E$,
acting over the extended space, together with a (technical) duality
transformation needed to preserve the locality of the interactions for
implementing properly the BP equations. The following two statements
will hold: {\it
\begin{itemize}
\item[{\bf (I)}] Marginals given by the BP equations derived from
$\hat E$ coincide with the marginals given by SP on the original
problem. 

\item[{\bf (II)}] Bethe approximation to the entropy of $\hat E$ in
the enlarged space as computed by BP coincides with the logarithm of
the number of clusters of solutions -- the so called ``complexity'' --
predicted by SP on the original problem.
\end{itemize}
} 

The proof of {\bf (I)} will be achieved by finding a direct connection
between quantities (``messages'') propagated by the two algorithms at
each iteration step.  We recall that the Bethe approximation to the
entropy is exact over trees without and with boundary conditions,
i.e. with leaf variables taking given values.

The possibility of interpreting SP as appropriate BP equations may
have consequences for their rigorous probabilistic analysis, through a
proper application/generalization of the known methods for the
analysis of convergence of BP like equations over random graphs (as it
has already been done for problems like the random matching
\cite{Aldous_z2}). Some preliminary exact numerical results that we
give in the concluding section are in support of this possibility.

Throughout the paper we heavily rely on the notations of
refs.~\cite{MZ,BMZ} for what concerns the SP equations.

\section{Survey Propagation, Belief Propagation and K-SAT}

SP and BP (or sum-product) are examples of message-passing procedures.
In BP the unknowns which are evaluated by iteration are the marginals
over the solution space of the variables characterizing the
combinatorial problem (e.g. binary ``spin'' variables). According to
the physical interpretation, the quantities that are evaluated by SP
are the probability distributions of local fields over the set of
clusters. That is, while BP performs a ``white'' average over
solutions, SP takes care of cluster to cluster fluctuations, telling
us which is the probability of picking up a cluster at random and
finding a given variable completely biased (frozen) in a certain
direction -- that is forced to take the same value within the cluster
-- or unfrozen.

In both SP or BP one assumes to know the marginals of all variables in
the temporary absence of one of them and then writes the marginal
probability induced on this ``cavity'' variable in absence of another
third variable interacting with it (i.e. the so called Bethe lattice
approximation for the problem).  These relations define a closed set of
equations for such cavity marginals that can be solved iteratively
(this fact is known as message-passing technique). The equations
become exact if the cavity variables acting as inputs are
uncorrelated. They are conjectured to be an asymptotically exact
approximation over some random locally tree--like structures\cite{MZ}.

The $K$-satisfiability problem ($K$-SAT) is easily stated: Given $N$
Boolean variables each of which can be assigned the value True (1) or
False (-1), and $M$ clauses between them, is there a 'SAT-assignment',
i.e. an assignment of the Boolean variables which satisfies all
constraints? A clause takes the form of an 'OR' function of $K$
variables in the ensemble (or their negations).  A SAT formula in
conjunctive normal form over $N$ Boolean variables $\{\sigma_i =\pm 1\}$ can
be written as
\begin{equation}
\mathcal{F}=\prod_{a\in A}C_{a}
\label{F}
\end{equation}
where 
\begin{equation}
C_a = 1 - E_a \; \; \; , \; \; E_a \equiv \prod_{i\in
a}\delta(J_{a,i},\sigma_i)
\label{eq:clause}
\end{equation}
where $\delta(x,y)$ is the Kronecker function (also written as
$\delta_{x,y}$ in the rest of the paper) and $\{ C_a \}$ are the
clauses encoded by the parameters $J_{a,i}$ as follows: $J_{a,i}=\pm
1$ if respectively $\pm \sigma_i$ appears in clause $a$ (in Boolean
notation we would have $J_{a,i}=-1$ (resp. $+1$) if the Boolean
variable $x_i$ (resp. $\neg x_i$) appears in clause $a$).  We call
$E_a$ the ``energy'' of a clause.  The symbol $i\in a$ will denote the
set of variables participating in clause $a$. Additionally it will be
useful to use the symbol $a\in i$ to denote the set of clauses
depending on variable $i$. The clause size $|\{i:i\in a\}|$ will be
denoted by $n_a$ ($n_a\equiv K$ for $K$-SAT), and the variable
connectivity $|\{a:a\in i\}|$ will be denoted by $n_i$.

The satisfiability problem consists in determining the existence of an
assignment to the Boolean variables which satisfies all clauses at the
same time, that is such that $\mathcal{F}=1$.  We may write the energy
function which counts the number of violated clauses as $E=\sum_a E_a$
so that the satisfiability problem becomes finding the zero energy
ground states of $E$. The random version of $K$-SAT corresponds to the
case in which the variables appearing in each clause are chosen
uniformly at random, and negated with probability $\frac12$. For the
sake of simplicity, hereafter we concentrate mostly on the $3$-SAT
case.

The energy function $E$ of a random $3$-SAT formula is a spin glass
model defined over a locally tree-like graph that can been studied
with the techniques of statistical physics of random systems, namely
the replica and cavity methods.

Numerical experiments have shown that a decimation algorithm based on
SP equations allows to find satisfying assignments of critically
constrained random $3$-SAT instances -- that is random formulas with
$\alpha=M/N$ just below a critical ratio $\alpha_c \simeq 4.267$ where
formulas are conjectured to become unsatisfiable with high probability
-- with a computational cost roughly scaling as $N \log N$~\cite{BMZ}
while the other known algorithms typically take times that are
exponential in $N$~\cite{Cook_review,nature}.  According to the cavity
-- or SP -- analysis , in such hard region (more precisely for $\alpha
\in [4.15,4.267]$~\cite{MZ,MPR}) there is a genuine one step RSB
phase, in which the space of solution decomposes into an exponential
number of clusters and where metastable states are even more numerous.

As discussed in great detail in ref.~\cite{MZ}, one crucial feature
that comes out from the SP analysis is the distinction between frozen
and unfrozen variables within the different clusters and we shall
introduce a formalism which naturally incorporates such phenomenon
(see also refs.~\cite{joker}).

We want to represent the condition for a variable of being not forced
to take any specific value in a given ground state (unfrozen) and to
this end we consider configuration space of $3-$value variables
$s_{i}\in\left\{-1,*,1,\right\} $ instead of $\sigma_i\in \{-1,1\}$.

We observe that $C_{a}$ as defined in Eq.~(\ref{eq:clause}) can be
evaluated also in extended variables: it behaves as if variables with
the $*$ value could be chosen to the best of $-1$ or $1$ and thus
satisfy the clause.  This gives the name ``joker state'' to the value
$*$. For a configuration $s^{(i,x)}$ such that $s^{(i,x)}_i = x$ and
$s^{(i,x)}_j = s_j$ for $j\neq i$ call
\begin{equation}
C_{a}^{i,x}(s)=C_{a}(s^{(i,x)})
\end{equation}
and introduce the constrain over $\{-1,*,1\}^{n}$ configurations given
by
\begin{equation}
V_{i}=\delta_{s_{i},*}\prod_{a\in i} C_{a}^{i,-1} C_{a}^{i,1} +
  \sum_{\sigma=\pm 1}\delta_{s_{i},\sigma} \prod_{a\in i}
  C_{a}^{i,\sigma}\left(1- \prod_{a\in i} C_{a}^{i,-\sigma}\right)
\label{eq:general2}
\end{equation}
The LEC formula derived from $\mathcal{F}$ will be defined as
\begin{equation}
\mathcal{G}=\prod_{i}V_{i}.  
\label{G}
\end{equation}
Note that $V_i$ depends only on $(s_{j})_{j\in a, a\in i}$ and
therefore preserves the ``locality'' of the structure, if any, of the
original formula.  A solution of the LEC problem is a configuration
$\mathbf{s}=(s_{i}) _{i\in I}\in\left\{-1,*,1\right\} ^{n}$ such that
$\mathcal{G}\left(\mathbf{s}\right)=1$. As a particular case, a
solution $\mathcal{G}(\mathbf{s})=1$ such that $s_{i}\in\left\{ \pm
1\right\} $ is also a solution of $\mathcal{F}$.

\begin{figure} 
\begin{center} 
\includegraphics[width=0.5\textwidth,height=0.3\textheight]{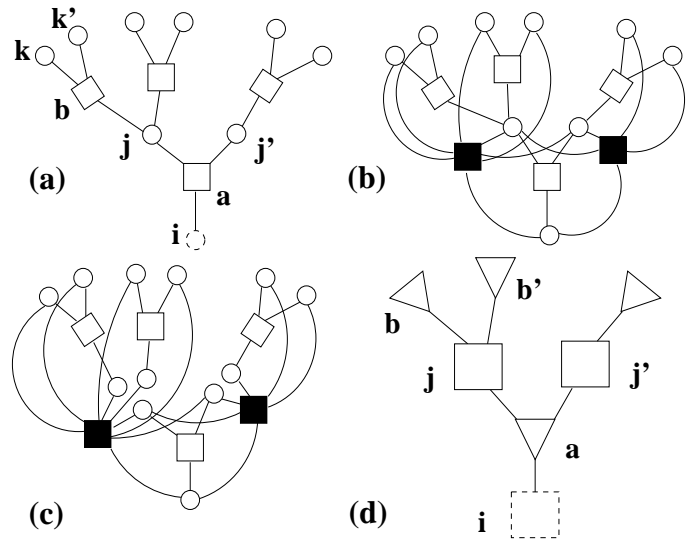} 
\caption{(a) Portion of the original factor graphs, (b) LEC graph with
3-state variables and additional constraints $A_i$ (black nodes) (c)
duality transformation (d) dual graph}
\label{duality}
\end{center} 
\end{figure}

To fix ideas it might be useful to compare the LEC cost-energy
function with the original 3-SAT one. To this end we adopt
the so--called factor graph representation~\cite{factor_graph}: Given
a formula $\mathcal{F}$, we define its associated \emph{factor graph}
as a bipartite undirected graph $G=\left(V;E\right)$, having two types
of nodes, and edges only between nodes of different type: {\bf (i)}
Variable nodes, each one labeled by a variable index in $I=\left\{
1,\dots,N\right\} $ and {\bf (ii)} Function nodes, each one labeled by
a clause index $a\in A$ ($|A|=M$). An edge $\left(a,i\right)$ will
belong to the graph if and only if $a\in i$ or equivalently $i\in a$.
For instance, the factor graph representation of the random $3$-SAT
problem consists in a bipartite graph with $N$ variable nodes having a
Poisson random connectivity of mean $3 \alpha$ and $M$ function nodes
with energy $E_a$ of uniform connectivity $3$ (a portion is shown in
part (a) of Fig.\ref{duality}).  The extended LEC spin glass energy
function reads:
\begin{equation}
\hat E =  \sum_{a=1}^M \hat E_a +\sum_{i=1}^N A_i 
\end{equation}
where now $\hat E_a = 1-C_a$ is evaluated in the extended configuration
space and
\begin{equation}
A_i=\delta_{s_{i},*}\left(1-\delta_{E_i^{-1},E_i^1}\right) +
\sum_{\sigma=\pm 1}\delta_{s_{i},\sigma}\theta\left(E_i^\sigma -
E_i^{-\sigma}\right)
\end{equation}
with $E_i^\sigma=\sum_{a\in i}(1-C^{i,\sigma}_a)$ and $\theta(x)=1$ if
$x > 0$ and $0$ otherwise.  The factor graph of the LEC has $N$
additional function nodes (the $A_i$ terms enforcing the joker
condition) that extend over the second neighbors (inset (b) in
Fig. \ref{duality}). 

By inspecting Eq.~(\ref{G}) we notice a first problem, namely that we
have lost the locally tree-likeness of the original graph. There are
interactions terms between every (ordered) pair of neighbors variable
nodes $i,j\in a$ (in the original graph), and thus for instance every
such pair shares two constraints $V_i,V_j$ (making an effective
2-loop). This introduces an obvious problem for implementing BP over
this combinatorial problem, and moreover would make difficult to
compare both algorithms, as the underlying geometry is now
different. Fortunately, there is an easy (but unfortunately
notationally somewhat involved) way out. We will group together
neighbor variables, effectively performing a sort of duality
transformation over the graph. We describe the procedure explicitly
below (Note that this is a particularly simple case of a Kikuchi or
``generalized belief propagation''-type approximation \cite{GBP}).

We will define: {\bf (i.)}  $M$ multi state variables each one
corresponding to a tuple $t_a=\{t^{(i)}_a\}_{i\in a}$ ($t^{(i)}_a\in
\{-1,*,1\}$) and ``centered'' on $a$ clauses and have (uniform)
connectivity $n_a$ ((c) in Fig.\ref{duality}), and {\bf (ii.)} N
function nodes $\chi^{dbp}_i$ having Poisson connectivity, depending
on $T_i\equiv \{t_a\}_{a\in i}$ and enforcing both the joker state
condition as well as identifying the values of the single variables
$t^{(i)}_a$ shared by different tuples $a\in i$ ((d) in
Fig.\ref{duality}).  An explicit expression of $\chi_i^{dbp}(T_i)$
(conf. Eq.~(\ref{eq:general2})) is
\begin{equation}
\chi^{dbp}_{i} = \sum_{\{s_i\}}\left(\prod_{a\in
i}\delta_{t_a^{(i)},s_i}\right) \left(\delta_{s_i,*}\prod_{a\in
i} C_{a}^{i,-1} C_{a}^{i,1} + \sum_{\sigma=\pm 1}\delta_{s_i,\sigma}
\prod_{a\in i} C_{a}^{i,\sigma}\left(1 - \prod_{a\in i}
C_{a}^{i,-\sigma}\right)\right)
\label{eq:dualenergy}
\end{equation}
We shall refer to the BP equations over the dual graph as {\it Dual
BP} (DBP).

\section{SP equations as BP equations over the dual graph}

Basic SP and DBP iterations can be thought of as transformations in
the space of probability distributions of the signs $h_i=\{-1,0,1\}$
of the effective fields acting on the single spin variables and of the
tuples $t_a=\{-1,*,1\}^{n_a}$ in the dual graph.  In the cavity
notation the quantities that are iterated refer to a graph in which a
given node and all its neighbor nodes are temporarily eliminated (see
Fig. \ref{duality} (a) and (d)) and all quantities are labeled by
oriented indices of the type $a \to i$ or $i \to a$ where the node on
the right of the arrow is the one eliminated.  Therefore the equations
describe a local transformation of some input probability
distributions into an output distribution in which a characteristic
function $\chi$ eliminates contributions from those combinations of
input and output fields or variables that violate some kind of local
constraints (it is worth noticing that these cavity equations are
closely related to the iterative local equations of the so called
Objective Method~\cite{Aldous} of combinatorial
probability). Explicitly we have:\\

{\bf DBP equations:}
\begin{eqnarray}
P_{a\to i}^{dbp}\left(t_a\right) & \propto & \sum_{\left\{
t_{b}\right\}} \prod_{j\in a\setminus i} \chi^{dbp}_j
\left(t_a,\left\{t_b\right\}\right) \prod_{b\in j\setminus a} P_{b\to
j}^{dbp}\left(t_{b}\right)
\end{eqnarray}
\\

{\bf SP equations:} ~\cite{MZ,BMZ}
\begin{eqnarray}
P_{j\to a}^{sp}\left(h_j\right) & \propto & \sum_{\left\{ h_k\right\}}
\chi^{sp}_{j\to a}\left(h_{j},\left\{h_{k}\right\}\right) \prod_{b\in
j\setminus a} \prod_{k\in b\setminus j} P^{sp}_{k\to
b}\left(h_{k}\right)
\end{eqnarray}
where 
\begin{eqnarray}
\chi^{sp}_{j\to a} & = & \delta_{h_{j},*}\prod_{b\in j\setminus
a}C_{b}^{j,1}C_{b}^{j,-1} + \sum_{\sigma=\pm
1}\delta_{h_{j},\sigma}\prod_{b\in j\setminus
a}C_{b}^{j,\sigma}\left(1-\prod_{b\in j\setminus
a}C_{b}^{j,-\sigma}\right)
\end{eqnarray}
$C_{b}$ clauses are here evaluated in $\left(\left(h_{k}\right)_{k\in
b\setminus j},h_{j}\right)$.\\

In order to show the connection between the above equations it is
convenient to introduce an auxiliary transformation $\tau$ of a
similar type:\\

{\bf $\tau$ transformation:}
\begin{eqnarray}
P_{a\to i}^{\tau}\left(t_a\right) & \propto & \sum_{\left\{
  h_j\right\}}\prod_{j\in a\setminus i} \chi^{\tau}_{j\to
  a}\left(t_a,h_j\right) P_{j\to a}\left(h_j\right)
\end{eqnarray}
and
\begin{eqnarray} 
\chi^{\tau}_{j\to a} = \sum_{\sigma=\pm 1} C_a \delta_{h_j,\sigma}
\delta_{t_a^{(j)},\sigma} + \delta_{h_j,*} \left[
\delta_{t_a^{(j)},*} C_{a}^{j,-1}C_{a}^{j,1} + \sum_{\sigma=\pm 1}
\delta_{t_a^{(j)},\sigma} C_{a}^{j,\sigma}\left(1 -
C_{a}^{j,-\sigma}\right)\right]
\label{eq:tau}
\end{eqnarray}
$C_{a}$ terms are evaluated here in $t_a$.\\

We will drop now the argument dependence of the measures $P_{j\to
a}^{sp}$, $P_{a\to i}^{dbp}$ and $P_{j\to a}^{\tau}$ and make instead
explicit the dependence on the input probability measures
$\left\{P_{k\to b}\right\},\left\{P_{b\to j}\right\},\left\{P_{j\to
a}\right\}$ respectively.

The connection between $DBP$ and $SP$ can be written as follows:
\begin{equation}
P_{a\to i}^{dbp}\left(\left\{P_{k\to b}^{\tau}\right\}\right) \equiv
  P_{a\to i}^{\tau}\left(\left\{P_{j\to a}^{sp}\right\}\right)
\label{eq:p_equiv}
\end{equation}
where both sides of the (functional) equality in turn depend on some
arbitrary set of probability distributions $\left\{P_k(h_k)\right\}$
where $k\in b\setminus j$ for $b\in j\setminus a$ and finally $j\in
a\setminus i$. In short,
\begin{equation}
P^{dbp}\circ P^{\tau}\equiv P^{\tau}\circ P^{sp}
\label{eq:equiv}
\end{equation}

In order to check the validity of the above identity we observe
that a direct inspection of the composition shows that it is true if
for every $j\in a\setminus i$ the following condition among the
characteristic functions holds:
\begin{equation}
\sum_{\{h_j\}}\chi^{\tau}_{j\to a} \chi_{j\to a}^{sp} =
\sum_{\{t_b\}}\chi^{dbp}_j\prod_{b\in
j\setminus a} \prod_{k\in b\setminus j}\chi^{\tau}_{k\to
b}\label{eq:compo}
\end{equation}
In appendix \ref{proof} we display the proof that this identity holds
and, as a consequence, that also identity Eq.~(\ref{eq:equiv}) is
valid. Eq.~(\ref{eq:equiv}) in turn implies that
\begin{equation}
\left(P^{dbp}\right)^{\left(k\right)}\circ P^{\tau}\equiv
P^{\tau}\circ\left(P^{sp}\right)^{\left(k\right)} \; \; \;,
\end{equation}
where the $\left(k\right)$ exponent means composition. This in turn
implies that we have a direct step-by-step connection between the
elementary quantities used in the DBP equations and those used in the
SP equations: convergence is obtained simultaneously and
Eq.~(\ref{eq:equiv}) holds for the respective fixed points.  It is
straightforward to compute from the $DBP$ equations the marginals
$P_{i}^{dbp}\left(s_{i}\right)$ of the single variables as a
marginalization of $P_{a}^{dbp}\left(t_{a}\right)$ for some $a\in i$
with respect to all other variables in the clause, (on a fixed point,
it doesn't matter which $a\in i$ one chooses). One finds that the
marginals predicted by DBP are in one to one correspondence with the
local fields given by SP, that is $P_i^{dbp}(s_i=-1,*,1)$ coincides
respectively with $P_i^{sp}(H_i=-1,0,1)$ (see refs.~\cite{MZ,BMZ}).

\subsection{Clustering and whitening}

The marginals over $\{1,*,-1 \}^N$ given by SP/DBP acquire a
computational/physical significance once we interpret what solutions
of combinatorial problem defined by Eq.~(\ref {G}) mean in term of
clusters (or groups) of solutions of the original problem defined by
Eq.~(\ref{F}). We will first define the Hamming distance between
configurations $s,t\in \{1,*,-1\}^n$, $H(s,t)=|\{i:s_i\neq t_i\}|$ and
an ordering relation over $\{-1,*,1\}$ configurations: if $s,t\in
\{1,*,-1\}^n$ we say that $s\leq t$ iff $t_i \neq s_i$ implies that
$t_i=*$. For instance, $(0,1)\leq(0,*)$ and $(1,1,1)\leq(1,*,*)$ but
$(0,1)\not\leq(1,*)$.

We will say that a configuration $s\in \{\pm 1\}^n$ is {\it contained}
in $t\in$ if $s\leq t$. In this sense, ``clustering'' would mean,
starting with some set $S\subset \{\pm 1\}^n$ of solutions of the
original combinatorial problem, to find some set $T \subset
\{1,*,-1\}^n$ such that every $s\in S$ is contained in some $t\in
T$. Of course, one would like to do so in some maximal way, but
satisfying some kind of separation between different clusters.

One trivial observation about the set ${\mathcal G}=1$ is that
solutions are by force separated, in the sense that $H(s,t) > 1$ if
${\mathcal G}(s)={\mathcal G}(t)=1$ and $s\neq t$. To prove this,
suppose that $H(s,t)=1$. If their difference comes because $s_i=\pm 1$
and $t_i=*$ then by force one of $V_i(t)$ or $V_i(s)$ is clearly
violated. If on the contrary, it comes because $s_i=1$ and $t_i=-1$ or
viceversa, then by force both of $V_i(t)$ and $V_i(s)$ are violated
and the only possible ``correct'' value for $s_i$ is $*$.

\begin{figure} 
\begin{center} 
\includegraphics[height=2cm]{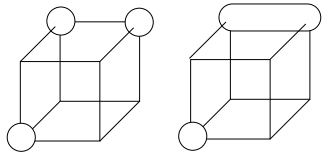}
\caption{The whitening procedure from left to right: the original set
of solutions $\{(-1,-1,-1), (1,1,-1), (1,1,1)\}$ and the set of
whitened clusters in the final step $\{(-1,-1,-1),(1,1,*)\}$}
\label{whitening-good}
\end{center} 
\end{figure}

A more important observation is that every solution of ${\mathcal
F}=1$ is {\it contained} in a solution of ${\mathcal G}=1$ with the
minimal number of $*$, and that solution can be easily found. Take a
solution $x$ of ${\mathcal F}=1$, and suppose that ${\mathcal G}=0$,
Choose a $V_i$ such that $V_i=0$. It can be easily seen that by
replacing $x_i$ by $*$, then $V_i$ becomes $1$. Then we pick another
violated constrain and repeat the process, until ${\mathcal G}=1$. We
will call the resulting configuration $w(x)$ (this procedure has been
already used under the name of {\it whitening} in the context of graph
coloring by G. Parisi in~\cite{joker}). It is easy to prove that the
result of this procedure does not depend on the order in which you
pick variables violating nodes $V_i$ (the proof being that any
violated $V_i$ will continue to be violated in the procedure, exactly
until we switch $x_i$ to $*$), and so $w(x)$ is uniquely defined. Note
that two configurations $x,y$ at Hamming distance $H(x,y)=1$ will have
$w(x)=w(y)$ and so every solution in a fixed connected component of
the solution space will end up inside the same ``cluster''. An example
of the whitening procedure for some set of solutions is depicted in
Figure~(\ref{whitening-good}).
An interesting point of view is that if one tries to build from
scratch a Hamiltonian to describe the behaviour of the outcomes of the
whitening procedure of some SAT formula, Eq.~(\ref{G}) comes
naturally.

\begin{figure} 
\begin{center} 
\includegraphics[height=2cm]{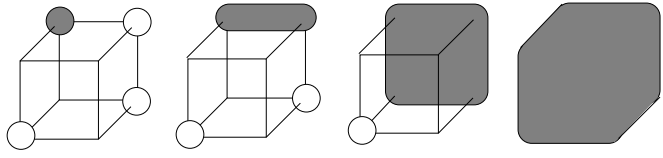}
\caption{A systematic error of the whitening $w((1,1,-1))$ (the dark
solution in the left). From left to right: the original sets of
solutions $\{(1,1,-1), (1,1,1), (1,-1,1), (-1,-1,-1)\}$ and first step
$(1,1,-1)$, second step $(1,1,*)$, third step $\{(1,*,*)\}$ and final
step $\{(*,*,*)\}$}
\label{whitening-errors}
\end{center} 
\end{figure}

The reader should note however that the presented definition of
clustering is far from perfect in the worst case: there is a number of
systematic errors produced by the whitening. For instance, in
Figure~(\ref{whitening-errors}) we can see one cluster claiming an
uncorrectly large volume. And there is of course also another problem:
unfortunately, there is no warranty that the sole solutions of
${\mathcal G}=1$ are the ones of the whitening, and in fact small
counter-examples can be easily constructed. Numerical work is being
done to ascertain a quantification of these two types of errors
(\cite{napolano}).

\section{Entropy and complexity}

The equivalence between the DBP marginals and the SP local field
probability distributions has the direct consequence that the Bethe
approximation to the entropy on the dual graph, $S^{dbp}$, coincides
with the logarithm of the number of clusters of solutions predicted by
SP, the so called complexity $\Sigma$.

On general grounds the Bethe approximation to the entropy of a problem
is exact if correlations among cavity variables can be neglected
(i.e. the global joint probability distribution takes a factorized
form). This is certainly true over tree graphs and it is conjectured
to be true in some cases for locally tree-like random graphs in the
limit of large size (one informal explanation is that distance between
cavity variables diverges with probability tending to one).
Factorization of marginal probabilities over our dual factor graph
amounts at writing $P(\{t_a\})=\prod_{i\in I} P^{dbp}_{i}(T_i)
\prod_{a \in A} [ P_a^{dbp}(t_a)]^{1-n_a}$ where $P_i^{dbp}(T_i)$ is
the joint probability distribution of the triples connected to node
$i$ ($T_i \equiv \{ t_b\}_{b \in i}$) and $P_a^{dbp}(t_a)$ is the
single triple marginal. Under this condition the entropy reads
\begin{eqnarray}
S = -\sum_{i}\sum_{\left\{ T_{i}\right\} }P^{dbp}_{i}(T_{i})\log
 P_{i}^{dbp}(T_{i}) + \sum_{a}\left(n_a - 1\right)\sum_{\left\{
 t_{a}\right\} }P^{dbp}_{a}(t_{a})\log P^{dbp}_{a}(t_{a}) \; .
\label{eq:entropy}
\end{eqnarray}

Showing $S=\Sigma$ is a straightforward calculation that we
report in the appendix. It requires to express the entropy in terms of
the cavity fields given by SP exploiting both Eq.~(\ref{eq:equiv}) and
the fixed point conditions. One finds
\begin{eqnarray}
S =  \sum_{i}\log c_{i}-\sum_{a}\left(n_{a} - 1\right)\log
 c_{a}-\sum_{i}\sum_{a\in i}\log D_{a\to i}
\label{eq:Sconst}
\end{eqnarray}
where the three normalization constants are defined by
\begin{eqnarray}
c_{i} & = & 
\sum_{\left\{ T_i\right\}}\prod_{a\in i}P_{a\to
i}\left(t_{a}\right)\chi_{i}\left(T_{i}\right)
\label{ci} \\
c_{a} & = & \sum_{t_a}\sum_{\left\{ h_{j}\right\}}\prod_{j\in
a}P_{j\to a}\left(h_{j}\right)\chi^{\tau}_{j\to
a}\left(h_{j},t_a\right)
\label{ca}\\
D_{a\to i} & = & \sum_{t_a}\sum_{\left\{ h_{j}\right\}}\prod_{j\in
a\setminus i}P_{j\to a}\left(h_{j}\right)\chi^{\tau}_{j\to
a}\left(h_{j},t_a\right)
\label{D}
\end{eqnarray}
These constants are not independent and the explicit expressions of
the first two are sufficient for writing $S$ in terms of SP
quantities:
\begin{eqnarray}
c_{a} & = & \sum_{\left\{ h_{j}\right\}} \prod_{j\in a} P_{j\to a}
 \left(h_{j}\right)\sum_{\left\{ t_a\right\}} \prod_{j\in a}
 \chi^{\tau}_{j\to a} \left(h_{j},t_a\right) \\ & = & 1 -
 \sum_{\left\{ h_{j}\right\}} \prod_{j\in a} P_{j\to a}
 \left(h_{j}\right)\left(1-\sum_{\left\{ t_{a}\right\}}\prod_{j\in a}
 \chi^{\tau}_{j\to a} \left(h_{j},t_a\right) \right) \\ & = & 1 -
 \prod_{j\in a}P_{j\to a}\left(J_{a,j}\right)\\ & = & 1- \prod_{j\in
 a} \frac{\Pi_{j\to a}^{u}}{\left(\Pi_{j\to a}^{s}+\Pi_{j\to
 a}^{0}+\Pi_{j\to a}^{u}\right)}
\end{eqnarray}
where we have borrowed the notation of Eq.~(18) in~\cite{BMZ}. For
computing $c_i$ we first notice that
\begin{equation} 
P_{a\to i}\left(t_{a}\right)=D_{a\to i}\sum_{\left\{ h_{j}\right\}
_{j\in a\setminus i}}\chi^{\tau}_{j\to
a}\left(t_{a},h_{j}\right)\prod_{j\in a\setminus i}P_{j\to
a}\left(h_{j}\right)
\end{equation}
so that Eq.~(\ref{ci}) reads
\begin{eqnarray} 
c_{i} & = & \prod_{a\in i} D_{a\to i}\sum_{\left\{ H_{i}\right\}
}\sum_{\left\{ T_{i}\right\} } \chi_{i} \left(T_{i}\right)
\prod_{a}\prod_{j\in a\setminus i}\chi^{\tau}_{j\to
a}\left(t_{a},h_{j}\right) P_{j\to a}\left(h_{j}\right)\nonumber \\ & = &
\prod_{a\in i}D_{a\to i}\sum_{\left\{ H_{i}\right\}
}\chi_i^{sp}(H_{i})\prod_{a}\prod_{j\in a\setminus i}P_{j\to
a}\left(h_{j}\right)\nonumber \\ & = & \prod_{a\in i}D_{a\to
i}\left(\hat{\Pi}_{i}^{+} + \hat{\Pi}_{i}^{0} +
\hat{\Pi}_{i}^{-}\right)
\end{eqnarray}
in the notations of Eq.~(21) in~\cite{BMZ}. Finally, plugging these
expressions into Eq.~(\ref{eq:Sconst}) and calling
\begin{eqnarray} w_{i} & = &
\hat{\Pi}_{i}^{+}+\hat{\Pi}_{i}^{0}+\hat{\Pi}_{i}^{-} \nonumber \\ x_{i\to a} & =
& \Pi_{j\to a}^{s}+\Pi_{j\to a}^{0}+\Pi_{j\to a}^{u}\nonumber \\ y_{i\to a} & =
& \Pi_{j\to a}^{u}
\end{eqnarray} 
we get from Eq. (\ref{eq:Sconst})
\begin{eqnarray}
S = \sum_{i}\log w_{i}- \left(n_{a} - 1\right) \sum_{a}
\log\left(1-\prod_{j\in a}\frac{y_{i\to a}}{x_{i\to a}}\right)
\label{esse1}
\end{eqnarray}
In this expression, $w_i$ represents the probability the local field
acting on the spin variable $i$ does not produce a contradiction and
$1 - \frac{y_{i\to a}}{x_{i\to a}}$ is the probability that the cavity
fields satisfy clause $a$.

We recall that the expression of the $SP$ complexity $\Sigma$ defined
in Eq.~(25-27) in~\cite{BMZ} is
\begin{eqnarray}
\Sigma & = & \sum_{i}\left(1-n_{i}\right)\log w_{i} +
 \sum_{a}\log\left(\prod_{i\in a}x_{i\to a} - \prod_{i\in a}y_{i\to
 a}\right) \nonumber \\ & = & \sum_{i}\log w_{i} - \sum_{a}\sum_{i\in a}\log
 w_{i} + \sum_{a}\log\left(\prod_{i\in a}x_{i\to
 a}-\prod_{i\in a}y_{i\to a}\right)
\label{sigma1} 
\end{eqnarray}
Despite their different look, it turns out that Eq.~(\ref{esse1}) and
Eq.~(\ref{sigma1}) are identical if evaluated in a fixed point of the SP
equations.  Their difference
\begin{eqnarray}
\label{eq:sigma-esse}
\Sigma - S = \sum_{a}\left\{ -\sum_{i\in a}\log w_{i} +
n_{a}\log\left(1-\prod_{i\in a}\frac{y_{i\to a}}{x_{i\to
a}}\right) - \sum_{i\in a}\log x_{i\to a}\right\}
\end{eqnarray}
is zero since in the fixed point every term inside the curly brackets
vanishes: using Eq.~(17) in~\cite{BMZ} we have that $\eta_{a\to
i}=\prod_{j\in a\setminus i}\frac{y_{j\to a}}{x_{j\to a}}$ ,
i.e. $\prod_{j\in a}\frac{y_{i\to a}}{x_{i\to a}}=\eta_{a\to
i}\frac{y_{i\to a}}{x_{i\to a}}$ for every $i\in a$ and hence
\begin{equation}
n_{a}\log\left(1-\prod_{j\in a}\frac{y_{i\to a}}{x_{i\to
a}}\right)=\sum_{j\in a}\log\left(1-\eta_{a\to i}\frac{y_{i\to
a}}{x_{i\to a}}\right)
\end{equation}
A simple calculation shows that $w_{i} = x_{a\to i}-\eta_{a\to
i}y_{a\to i}$ for every $a\in i$ and therefore we get $\Sigma=S$ as
desired.

\section{Discussion and Conclusions}

In this work we have shown by elementary means that the SP equations
can be interpreted and derived as sum-product equations for the
marginals over a modified combinatorial problem. An important
consequence of this fact is a clarification of the hypothesis behind
the algorithm. It is to be expected that the essential hypothesis
making sum-product to work is the uncorrelation of the marginals of
distant (or cavity) variables. Under the shown mapping, this directly
implies that the hypothesis behind SP (and in a way, of its definition
of clusters) is the uncorrelation of the frozen part of distant
variables, that is the uncorrelation between {\bf different} clusters.

Under this light one can think of the SP procedure of obtaining $\hat
E$ from $E$ as a way of collapsing the internal structure of pure
states: the resulting problem ${\mathcal G}$ has many pure states but
with zero internal entropy. Note that this is a completely different
limit case with respect to the ``one pure state''  in which BP
(more precisely DBP) is shown to work correctly and to predict an
accurate entropy (which we remind is the complexity of the original
$E$).

As far as the connection between solutions of the modified problem and
the original one is concerned, things are particularly simple over
tree factor graphs (see also \cite{BMZ} for results concerning
propagation of messages): Indeed, for any fixed boundary condition
(i.e. an assignment for the leaf variables), there is at most one
solution with $\hat{E} = 0$, and it is easy to prove (see
appendix~\ref{tree}) that all solutions of $E=0$ correspond to the
same connected component of the solution space (i.e. every two
solutions can be joined by a path of solutions in which successive
configurations in the path differ by exactly one spin flip). 

The situation on loopy graphs (corresponding for instance to random
formulae) is obviously more complicated. A coherent interpretation
would be that not only the recursive $DBP$/$SP$ equations themselves
are accurate in a probabilistic sense (i.e. when the factorization of
the corresponding input joint probability is sound) to compute the
statistics of the ground states of $\hat E$, but also that the
exactness of the interpretation of the ground states of $\hat E$ in
terms of clustering of the ground states of $E$ relies on this
hypothesis being true.

To this extent we mention that exact enumerations on a large number
(thousands) of small random 3-sat formulas (up to $N=100$) showed that
all the zero energy configurations of $\hat E$ which are stable under
SP iterations can be extended to real solution of the original
problem.  Spurious ground states (i.e. configurations that are not
extensible to real solutions) do exist with a non negligible
probability for small $N$, however they turn out to be always unstable
fixed points of SP , that is unsat configurations which are irrelevant
for the SP marginals \cite{napolano}.  While such a result was
expected to hold for tree-like graphs, it is somewhat surprising to
observe it numerically on small, loopy, random factor graphs.  The
robustness of such result calls for a finite $N$ probabilistic analysis
which would represent a building brick for the rigorous analysis of SP
(of course, small ad-hoc counterexamples on improbable formulae can be
easily constructed).

As a concluding remark we notice that the discussed formalism can be
generalized to take care of the non-zero energy regime where not all
constraints can be satisfied simultaneously (``frustrated'' case). The
LEC energy function takes the form $\hat E= \lambda \sum_{a\in A} \hat
E_a + \sum_{i\in I} A_i$, where $\lambda$~\cite{note} plays the role
of the so called Parisi re-weighting parameter~\cite{cavity}. \\

\section{Acknowledgments}

We thank D. Achlioptas, M. Mezard, G. Parisi, A. Pelizzola and M.
Pretti for very fruitful discussions. This work has been supported in
part by the European Community's Human Potential Programme under
contract HPRN-CT-2002-00319, STIPCO.

\appendix

\section{Proof of equivalence}

\label{proof}

For the LHS of Eq.~(\ref{eq:compo}) we have:\\

\noindent
If $h_j = \sigma\in\{\pm 1\}$ then
\begin{equation}
 \chi^{\tau}_{j\to a}= C_a\delta_{t_a^{(j)},\sigma} \; \; \; , \; \; 
 \chi^{sp}_{j\to a} = \prod_{b\in
j\setminus a}C_b \left(1-\prod_{b\in j\setminus a}
C_{b}^{j,-\sigma}\right)
\end{equation}
\noindent
If $h_j = *$ then
\begin{equation}
\chi^{\tau}_{j\to a} = \delta_{t^{(j)}_a,*}C_{a}^{j,-1}C_{a}^{j,1} +
\sum_{\sigma=\pm 1}\delta_{t_a^{(j)},\sigma} C_{a}^{j,\sigma}\left(1 -
C_{a}^{j,-\sigma}\right) \; \; \; , \; \; \chi^{sp}_{j\to a} =
\prod_{b\in j\setminus a}C_{b}^{j,-1}C_{b}^{j,1}.
\end{equation}

Summing up both products and regrouping the LHS of Eq.~(\ref{eq:compo}) reads:
\begin{eqnarray}
\sum_{\sigma=\pm 1}\delta_{t_a^{(j)},\sigma} \prod_{b\in j}
C_{b}^{j,\sigma} \left(1 - \prod_{b\in j} C_{b}^{j,-\sigma}\right) +
\delta_{t_a^{(j)},*} \prod_{b\in j} C_{b}^{j,-1}
C_{b}^{j,1}\label{eq:sptau}
\end{eqnarray}
where $C_{b}$ for $b\in j\setminus a$ is evaluated here in
$\left(\{h_k\}_{k\in b\setminus j},t_a^{(j)}\right)$ and $C_{a}$ is
evaluated in $t_a$.

For the RHS of Eq.~(\ref{eq:compo}) we first notice that as the
$\chi^{dbp}_{j}$ term includes $\prod_{a\in j} \delta_{t_a^{(j)},s_j}$
we will simply replace all occurrences of $t_b^{(j)}$ and $s_j$
variables by $t_a^{(j)}$ and drop the outer sum and the product term
itself. For instance, the sum over $\{ t_b\}_{b\in j}$ thus reduces to
a sum over $\left\{\{ t_b^{(k)}\}_{k\in b\setminus
j},{t_a^{(j)}}\right\}$. Let's evaluate the RHS of
Eq.~(\ref{eq:compo}) on the three possible values of $t_a^{(j)}$:\\
\noindent
If $t_a^{(j)} = *$ then by Eq.~(\ref{eq:dualenergy}) $\chi^{dbp}_j =
\prod_{b\in j}C_{b}^{j,-1}C_{b}^{j,1}$.  Moreover, just by looking at
its definition Eq.~(\ref{eq:tau}), one finds that in
$\chi^{\tau}_{k\to b}$ all $C$ terms are equal to $1$ since their $j$
coordinate $t_b^{(j)}=t_a^{(j)}$ is $*$.  Then $\chi^{\tau}_{k\to b} =
\delta_{t_b^{(k)},h_k}$ and the RHS of Eq.~(\ref{eq:compo}) becomes
\begin{equation}
C_{a}^{j,-1} C_{a}^{j,1} \prod_{b\in j\setminus a} C_{b}^{j,-1}
C_{b}^{j,1} \prod_{k\in b\setminus j} \delta_{t_b^{(k)},h_k}
\end{equation}
which is exactly the term in Eq.~(\ref{eq:sptau}) corresponding to
$t_a^{(j)} = *$ (remember that $C_{b}$ clauses here are evaluated in
$t_b$).\\
\noindent
If $t_a^{(j)} = \sigma\in\{\pm 1\}$ then it is convenient to break
$\chi^{dbp}_j$ in two addenda:
\begin{equation}
\prod_{b\in j} C_{b}-\prod_{b\in j} C_{b} C_{b}^{j,-\sigma}
\end{equation}
so that the RHS of Eq.~(\ref{eq:compo}) becomes
\begin{eqnarray}
C_{a}\prod_{b\in j\setminus a} \left(\sum_{\{t_{b}\}}C_{b}\prod_{k\in
  b\setminus j}{\chi^{\tau}_{k\to b}}\right) - C_{a}C_{a}^{j,-\sigma}
  \prod_{b\in j\setminus
  a}\left(\sum_{\{t_{b}\}}C_{b}C_{b}^{j,-\sigma}\prod_{k\in b\setminus
  j}\chi^{\tau}_{k\to b}\right)
\end{eqnarray}
Finally, both sums can be computed explicitly and the result is
again exactly the corresponding term in Eq.~(\ref{eq:sptau}). This ends
the proof of the identity Eq.~(\ref{eq:equiv}).

\section{Computation of the entropy}
\label{entropy-complexity}
For simplicity of notation, in what follows we write $P_{a}(t_{a}),
P_{a\to i}(t_a), P_i(T_i)$ and $\chi_i(T_i)$ in place of
$P_{a}^{dbp}(t_{a}), P_{a\to i}^{dbp}(t_a), P_i^{dbp}(T_i)$ and
$\chi_i^{dbp}(T_i)$ respectively and $P_{i\to a}(h_i)$ in place of
$P_{i\to a}^{sp}(h_i)$.

To compute the entropy (\ref{eq:entropy}) we first need
\begin{eqnarray*}
P_{a}(t_{a}) & = & c_{a}^{-1}\sum_{\left\{ h_i \right\}} \prod_{i\in
 a}P_{i\to a}\left(h_i\right)\prod_{i\in a}\chi^{\tau}_{i\to
 a}\left(t_{a},h_i\right)\\ & = & c_{a}^{-1}\prod_{i\in
 a}\sum_{\left\{ h_i\right\}} P_{i\to
 a}\left(h_i\right)\chi^{\tau}_{i\to a}\left(t_{a},h_i\right)
\end{eqnarray*} 
Thus calling
\begin{equation}
f_{a\to i}=\sum_{\left\{ h_i\right\} }P_{i\to
a}\left(h_i\right)\chi^{\tau}_{i\to a}\left(t_{a},h_i\right)
\end{equation}
we have that
\begin{eqnarray}
\sum_{\left\{ t_{a}\right\} }P_{a}(t_{a})\log P_{a}(t_{a}) =
-c_{a}^{-1}\log c_{a}+\sum_{\left\{ t_{a}\right\}
}P_{a}(t_{a})\sum_{i\in a}\log f_{a\to i}\nonumber \\ =
-c_{a}^{-1}\log c_{a}+\sum_{i\in a}\sum_{\left\{ t_{a}\right\}
}P_{a}(t_{a})\log f_{a\to i}\label{eq:ca}
\end{eqnarray}
Writing  $\omega_{a\to i}=\sum_{\left\{ t_{a}\right\} }P_{a}(t_{a})\log
f_{a\to i}$ we get
\begin{eqnarray}
\sum_{a}\left(n_{a} - 1\right)\sum_{i\in a}\omega_{a\to i} &=&
\sum_{i}\sum_{a\in i}\sum_{j\in a\setminus i}\omega_{a\to j}\nonumber
\\ &=& \sum_{i}\sum_{a\in i}\sum_{j\in a\setminus i}\sum_{\left\{
t_{a}\right\} }P_{a}(t_{a})\log f_{a\to j}\nonumber \\
 &=& \sum_{i}\sum_{a\in i}\sum_{\left\{ t_{a}\right\} }
P_{a}(t_{a})\prod_{j\in a\setminus i}\log f_{a\to j}\nonumber \\
 &=& \sum_{i}\sum_{a\in i}\sum_{\left\{ t_{a}\right\} }\sum_{\left\{
t_{b}\right\} _{b\in i\setminus a}}P_{i}(T_{i})\prod_{j\in a\setminus
i}\log f_{a\to j}\nonumber \\  &=& \sum_{i}\sum_{a\in i}\sum_{\left\{
T_{i}\right\} } P_{i}(T_{i}) \log\prod_{j\in a\setminus i}f_{a\to
j}\label{eq:star}
\end{eqnarray} 
The term inside the logarithm above reads
\begin{eqnarray} 
\prod_{j\in a\setminus i}f_{a\to j} =  \sum_{\left\{ h_j\right\}}
 \prod_{j\in a\setminus i} \chi^{sp}_{j\to a} \left(t_{a},h_j\right)
 \prod_{j\in a\setminus i} P_{j\to a}(h_j)  =  \frac{1}{D_{a\to
 i}}P_{a\to i}(t_{a})
\end{eqnarray}
where $D_{a\to i}$ is an appropriate normalization constant. Going
back to Eq.~(\ref{eq:star}), we have
\begin{eqnarray}
\sum_{a}(n_{a}-1)\sum_{i\in a}\omega_{a\to i} = -\sum_{i}\sum_{a\in
i}\log D_{a\to i} + \sum_{i}\sum_{a\in i}\sum_{\left\{ T_{i}\right\}
}P_{i}\left(T_{i}\right)\log P_{a\to i}(t_{a})
\label{eq:D+P}
\end{eqnarray} 
The second term in the right-hand side equals
\begin{eqnarray}
\sum_{i}\sum_{\left\{ T_{i}\right\}
 }P_{i}\left(T_{i}\right)\log\prod_{a\in i}P_{a\to i}(t_{a})
  & = & \sum_{i}\sum_{\left\{ T_{i}\right\}
 }P_{i}\left(T_{i}\right)\log\chi_{i}(T_{i})\prod_{a\in i}P_{a\to
 i}(t_{a})\nonumber \\ & = & \sum_{i}\sum_{\left\{ T_{i}\right\}
 }P_{i}\left(T_{i}\right)\log Q_{i}(T_{i})\nonumber \\ & = &
 \sum_{i}\sum_{\left\{ T_{i}\right\} }P_{i}\left(T_{i}\right)\log
 P_{i}(T_{i})+\sum_{i}\sum_{\left\{ T_{i}\right\}
 }P_{i}\left(T_{i}\right)\log c_{i}
\label{eq:P}
\end{eqnarray}
where in the second step above $\chi_{i}(T_{i})$ has been artificially
multiplied inside the logarithm (we can do it because there is a
$P_{i}(T_{i})$ outside) and $P_{i}(T_{i}) =
\frac{1}{c_{i}}Q_{i}(T_{i})$. Eqs.~(\ref{eq:D+P}),(\ref{eq:P}) give:
\begin{eqnarray} 
\sum_{a}(n_{a}-1)\sum_{i\in a}\omega_{a\to i} = -\sum_{i}\sum_{a\in
  i}\log D_{a\to i} +\sum_{i}\sum_{\left\{ T_{i}\right\}
  }P_{i}\left(T_{i}\right)\log P_{i}(T_{i})+\sum_{i}\log
  c_{i}
\label{eq:D}
\end{eqnarray}
Going back to the first expression of the entropy
Eq.~(\ref{eq:entropy}), and using Eq.~(\ref{eq:ca}) and
Eq.~(\ref{eq:D}) we get:
\begin{eqnarray}
S & = & -\sum_{i}\sum_{\left\{ T_{i}\right\} } P_{i}(T_{i})\log
 P_{i}(T_{i}) + \sum_{a}\left(n_{a} - 1\right)\sum_{\left\{
 t_{a}\right\} }P_{a}(t_{a})\log P_{a}(t_{a})\nonumber \\ & = &
 \sum_{i}\log c_{i}-\sum_{i}\sum_{\left\{ T_{i}\right\}
 }P_{i}(T_{i})\log Q_{i}(T_{i}) +
 \sum_{a}\left(n_{a}-1\right)\sum_{\left\{ t_{a}\right\}
 }P_{a}\left(t_{a}\right)\log P_{a}(t_{a})\nonumber \\ & = &
 \sum_{i}\log c_{i}-\sum_{a}\left(n_{a}-1\right)\log
 c_{a}-\sum_{i}\sum_{a\in i}\log D_{a\to i}
\end{eqnarray}
where the constants are defined in Eqs.~(\ref{ci}-\ref{D}).

\section{Tree factor graphs}
\label{tree} 
The argument turns out to be similar to the one given in an analogous
``tutorial'' appendix in ref.  \cite{Barthel_Hartmann} for the Vertex
Cover problem.\\ 
We will first build a reference solution ${\mathbf x}$, and then show
that every solution of $E=0$ is connected to it. ${\mathbf x}$ will be
built from the leaves to the root. Suppose the variables are labeled
in an ordering that respects distances to the root, such that the
first ones are the leaves and the last one is the root. In such an
ordering, the parents (resp. child) of $i$ are neighbors with labels
$j<i$ (resp.  $j>i$). We will fix $x_i$ iteratively: once $x_j$ for
$j<i$ are fixed, all parents of $j$ are fixed; then for $x_j$ there
are two possibilities: either its parents force it to take a specific
value, or they don't. In the first case we chose $x_i$ to take the
forced value; in the second one we chose the value that satisfy the
child clause. Now we can show that ${\mathbf x}$ is connected with
every other solution ${\mathbf s}$ (and thus every two solution are
connected). It is easy to see that the configurations ${\mathbf
y}^{(k)}$ defined by ${\mathbf y}^{(k)}_j = s_j$ if $j<k$ and
${\mathbf y}^{(k)}_j = x_j$ if $j\geq k$ form a path of configurations
connecting ${\mathbf x}$ and ${\mathbf s}$. Clearly ${\mathbf
y}^{(1)}={\mathbf x}$ and ${\mathbf y}^{(n)}={\mathbf s}$. Also they
are all solutions, since if ${ \mathbf y}^{(k)}$ is a solution, then
clearly ${\mathbf y}^{(k+1)}$ is also a solution: if they are
different it is because ${\mathbf y}^{(k+1)}_{k+1}$ has been chosen to
satisfy the child clause (and it was not forced from parents in $s$
and thus neither in $y^{(k+1)}$).

We can now look for solutions of $\hat E$ on a satisfiable tree (with
boundary conditions). Let's start with a free-boundary tree with $2$
and $3$-clauses: it is easy to see that the solution with all $*$
assignments has $\hat E = 0$. It is also clearly unique: suppose that
there is a solution with some variable set to $\sigma\neq *$. Then
there is forcefully one of its neighboring clauses in which the two
(or one) remaining variables are fixed in order to not satisfy the
clause. Repeating again the argument recursively for one of them, we
can get a never-ending path of fixed variables in the tree. But as a
trees have no loops, this is a contradiction.

There is also exactly one such solutions for a satisfiable tree with
boundary conditions (if we disregard $V_i$ constraints on the
variables with assigned boundary values). We will build it explicitly
using the so-called unit clause propagation (UCP).  The UCP procedure
consists in removing (in this case starting from the boundary) every
fixed variable by (a) removing all clauses satisfied by the variable
and (b) removing the variable from all clauses in which it appears
without satisfying the clause.  (if the original tree is satisfiable,
no $0$-clause can appear in this erasure step).  Then every possibly
appearing $1$-clause is taken and its variable fixed in order to
satisfy the clause, and the procedure starts again from the beginning
until no more $1$-clauses show up. The resulting graph is
boundary-free and with no $1$-clauses.

The promised solution will be built by taking all variables fixed by
UCP with their assigned value, and by assigning the value $*$ to the
remaining ones. The resulting configuration $\hat x$ has $\hat E(\hat
x)=0$.  Clearly the constraints $V_i$ (see Eq.~(\ref{eq:general2}))
are satisfied by $\hat x$ for all $i$ fixed by UCP (because they are
``frozen'' by their neighbors). We easily see that this partial
assignement is the unique one that can give $\hat E = 0$. Using the
fact that the subgraph produced by UCP has no boundary condition and
that the unique solution for $\hat E=0$ on that subgraph is the
all-$*$ one, we see that the proposed configuration is indeed the
unique solution.

Note also that every solution of $E=0$ will coincide with $\hat x$ in
the $-1,1$-assigned variables of the latter, because these variables
were fixed by UCP and thus are forced in every satisfying
configuration. Moreover, if one takes an index $i$ such that $\hat
x_i$ is $*$, then there is at least one solution of $E(s)=0$ with
$s_i=1$ (resp. $-1$): by fixing $s_i$ and applying again UCP one
cannot get any contradiction (i.e. a $0$-clause) because the subgraph
has no loops nor $1$-clauses. The remaining graph is still loop-free,
and thus trivially satisfiable.


\begin{thebibliography}{99}

\bibitem{TCS} Special Issue on {\it NP-hardness and Phase transitions},
O. Dubois, R. Monasson, B. Selman and R. Zecchina (eds.),
Theor. Comp. Sci. \textbf{265}, Issue: 1-2, August 28 (2001).

\bibitem{Codes} H. Nishimori, {\it Statistical Physics of Spin Glasses and
Information Processing}, Oxford University Press, 2001

\bibitem{Aldous} D. Aldous, J. M. Steele, Probability on Discrete
Structures (Vol. 110 of Encyclopaedia of Mathematical Sciences),
ed. H. Kesten, p. 1-72. Springer, 2003.

\bibitem{Guerra:Talagrand} F. Guerra, Comm. Math. Phys. {\bf
233}, 1 (2003); M. Talagrand, C.R. Acad. Sci. Paris, Ser. I {\bf 337},
111 (2003)

\bibitem{Aldous_z2} D. Aldous, Random Structures and Algorithms {\bf
18} 381 (2001)


\bibitem{MPV} M. Mezard, G. Parisi, M.A. Virasoro, {\it Spin Glass
Theory and Beyond}, World Scientific, (1987)

\bibitem{pspin} S. Cocco, O. Dubois, J. Mandler, R. Monasson.
Phys. Rev. Lett. {\bf 90}, 047205 (2003); M. Mezard, F. Ricci-Tersenghi,
R. Zecchina, J. Stat. Phys. {\bf 111}, 505 (2003)

\bibitem{science} M. Mezard, G. Parisi, R. Zecchina, Science {\bf
297}, 812 (2002)

\bibitem{MZ} M. Mezard and R. Zecchina, Phys.Rev. {\bf E 66}, 056126 (2002)

\bibitem{BMZ} A. Braunstein, M. Mezard, R. Zecchina, {\it Survey
propagation: an algorithm for satisfiability}, ArXiv:
xxx.lanl.gov/ps/cs.CC/0212002 (2002)

\bibitem{Gallager} R.G. Gallager, Information Theory and Reliable
  Communications, Wiley, New York, 1968

\bibitem{Pearl} J. Pearl, {\sl Probabilistic Reasoning in Intelligent Systems},
2nd ed. (San Francisco, MorganKaufmann,1988)

\bibitem{Spielman} D.A. Spielman, in {\sl Lecture Notes in Computer
  Science} {\bf 1279}, 67 (1997) 

\bibitem{Sourlas} N. Sourlas, in {\sl From Statistical Physics to
Statistical Inference and Back}, P. Grassberger and J-P. Nadal Edts.,
Kluwer Academic, Dordrecht (1994)

\bibitem{turbo} C. Berrou, A. Glavieux and P. Thitimajshima,
  Proc. Int. Conf. Comm, 1064-1070 (1993)

\bibitem{Forney} G.D. Forney, Jr., IEEE Trans. Inform. Theory, {\bf
  47}, 520 (2001)

\bibitem{good_codes1} M.G. Luby, M. Mitzenmacher, M.A. Shokrollahi and
  D.A. Spielman, IEEE Trans. Inform. Theory, {\bf 47}, 569 (2001)

\bibitem{good_codes2} S-Y. Chung, G.D. Forney,Jr., T.J. Richardson and
  R. Urbanke, IEEE Comm. Letters {\bf 5}, 58 (2001)

\bibitem{MacKay} D.J.C. MacKay, IEEE Trans. Inform. Theory {\bf 45},
  399 (1999)

\bibitem{cavity} M. Mezard, G. Parisi, M.A. Virasoro, Europhys. Lett. {\bf
1}, 77 (1986); M. Mezard, G. Parisi, Eur. Phys. J. {\bf B 20}, 217
(2001); M. Mezard, G. Parisi, J. Stat. Phys.  {\bf 111}, 1 (2003)


\bibitem{Cook_review} S.A. Cook, D.G. Mitchell, {\it Finding Hard
Instances of the Satisfiability Problem: A Survey}, In: {\sl
Satisfiability Problem: Theory and Applications}, Du, Gu and Pardalos
(Eds).  DIMACS Series in Discrete Mathematics and Theoretical Computer
Science, Volume 35, (1997)

\bibitem{nature} R. Monasson, R. Zecchina, S. Kirkpatrick,
B. Selman, and L.  Troyansky, Nature \textbf{400}, 133 (1999);

\bibitem{MPR} A. Montanari, G. Parisi, F. Ricci-Tersenghi, ArXiv:
xxx.lanl.gov/ps/cond-mat/0308147 (2003)

\bibitem{joker} 
A. Braunstein, M. Mezard, M. Weigt, R. Zecchina, {\it Constraint
Satisfaction by Survey Propagation}, ArXiv
lanl.arXiv.org/ps/cond-mat/0212451 (2002);
G. Parisi, {\it On the survey-propagation equations for the random
K-satisfiability problem}, ArXiv: xxx.lanl.gov/ps/cs.CC/0212009
(2002);
G. Parisi, {\it On local equilibrium equations for clustering states}
ArXiv: xxx.lanl.gov/ps/cs.CC/0212047 (2002);

\bibitem{factor_graph} F.R. Kschischang, B.J. Frey, H.-A. Loeliger,
{\it IEEE Trans. Infor. Theory} {\bf 47}, 498 (2002).

\bibitem{GBP} Yedidia, J.S.; Freeman, W.T.; Weiss, Y., {\it Generalized
Belief Propagation}, Advances in Neural Information Processing Systems
(NIPS) {\bf 13}, 689 (2000)

\bibitem{states} There exist multiple definitions of states (clusters)
for finite sizes (e.g.  k-flip stable, with $lim_{N \to \infty} k/N
=0$, \cite{cavity,Biroli:Monasson}) which lead to equivalent
thermodynamical limits in which the SP-cavity formalism is assumed to
hold.

\bibitem{pspin-note} In particularly simple cases like the so called
diluted p-spin glasses (or random sparse parity check
equations)~\cite{pspin}, the introduction of $*$ states has allowed
for an explicit construction of an exponential number of clusters of
solutions and to prove the exactness of the so called one step replica
symmetry breaking (RSB) solution in the scheme of Parisi~\cite{MPV}.
However, for such models the $*$ variables are in a sense trivial in
that they do not depend on the cluster and their (recursive)
elimination leads to a residual model which can be solved exactly by a
simple annealed/first-moment calculation.  For K-SAT the situation is
more complex (and more general) in that variables are expected to
become $*$ depending on the clusters.


\bibitem{note} In the computation of the free energy $\lambda$ should
be taken proportional to the temperature $T$ in the limit $T \to 0$.

\bibitem{Biroli:Monasson} G. Biroli, R. Monasson, Europhys. Lett. 50,
155 (2000)


\bibitem{Barthel_Hartmann} W. Barthel, A.K. Hartmann,{\it Clustering
analysis of the ground-state structure of the vertex cover problem},
cond-mat/0403193

\bibitem{napolano} A. Braunstein, V. Napolano, R. Zecchina {\it
Clustering in random SAT}, in preparation


\end{thebibliography}
\end{document}